\documentclass[sigconf]{acmart}

\usepackage{xcolor,colortbl}
\usepackage{graphicx}
\usepackage{bm}
\usepackage{enumitem}
\usepackage[utf8]{inputenc}
\usepackage[english]{babel}
\usepackage{multirow}
\usepackage{makecell}
\usepackage{booktabs} 
\usepackage{subcaption}
\usepackage{amsmath}
\theoremstyle{definition}
\newtheorem{definition}{Definition}
\newtheorem*{assumption}{Assumption}
\newtheorem*{proposition}{Proposition}

\usepackage{footnote}
\makesavenoteenv{tabular}
\usepackage[font=small]{caption}
\usepackage{dsfont}
\newcommand{\mat}[1]{{\bm #1}}   

\AtBeginDocument{%
  \providecommand\BibTeX{{%
    \normalfont B\kern-0.5em{\scshape i\kern-0.25em b}\kern-0.8em\TeX}}}

\setcopyright{acmcopyright}
\copyrightyear{2021}
\acmYear{2021}
\setcopyright{acmcopyright}\acmConference[KDD '21]{Proceedings of the 27th ACM SIGKDD Conference on Knowledge Discovery and Data Mining}{August 14--18, 2021}{Virtual Event, Singapore}
\acmBooktitle{Proceedings of the 27th ACM SIGKDD Conference on Knowledge Discovery and Data Mining (KDD '21), August 14--18, 2021, Virtual Event, Singapore}
\acmPrice{15.00}
\acmDOI{10.1145/3447548.3467321}
\acmISBN{978-1-4503-8332-5/21/08}

\begin{document}
\fancyhead{}
\title[Causal Understanding of Fake News Dissemination on Social Media]{Causal Understanding of Fake News Dissemination \\on Social Media}

\author{Lu Cheng\textsuperscript{\rm 1},  Ruocheng Guo\textsuperscript{\rm 1}, Kai Shu\textsuperscript{\rm 2}, Huan Liu\textsuperscript{\rm 1}}
\affiliation{\textsuperscript{\rm 1} Computer Science and Engineering, Arizona State University, USA\\
\textsuperscript{\rm 2}  Department of Computer Science, Illinois Institute of Technology, USA}
\email{{lcheng35, rguo12, huanliu}@asu.edu, kshu@iit.edu}

\renewcommand{\shortauthors}{Cheng et al.}

\begin{abstract}
Recent years have witnessed remarkable progress towards computational fake news detection. To mitigate its negative impact, we argue that it is critical to understand what user attributes potentially \textit{cause} users to share fake news. The key to this causal-inference problem is to identify \textit{confounders} -- variables that cause spurious associations between treatments (e.g., user attributes) and outcome (e.g., user susceptibility). In fake news dissemination, confounders can be characterized by fake news sharing behavior that inherently relates to user attributes and online activities. Learning such user behavior is typically subject to \textit{selection bias} in users who are susceptible to share news on social media. Drawing on causal inference theories, we first propose a principled approach to alleviating selection bias in fake news dissemination. We then consider the learned \textit{unbiased} fake news sharing behavior as the surrogate confounder that can fully capture the causal links between user attributes and user susceptibility. We theoretically and empirically characterize the effectiveness of the proposed approach and find that it could be useful in protecting society from the perils of fake news.  
\end{abstract}




\maketitle
\section{Introduction}
\label{sec1}
Online social media ushers the world to an unprecedented time of ``fake news'' -- false or misleading information disguised in news articles to mislead consumers
\cite{guess2019less,shu2017fake}. This has raised serious concerns, demanding novel approaches to understanding fake news dissemination. While great effort can be seen in computational fake news detection, less is known about what user attributes \textit{cause} some users to share fake news. In contrast to the research focused on correlations between user profiles (e.g., age, gender) and fake news (e.g., \cite{shu2018understanding}), this work seeks a more nuanced understanding of how user profile attributes are \textit{causally} related to user susceptibility to share fake news\footnote{As we cannot know the exact intentions of users who spread fake news (e.g., gullible or malicious users) using only observed user engagement data, we propose a measure to approximate user susceptibility as detailed in Sec. 4.3.}. The key to identifying causal user attributes with observational data is to find \textit{confounders} -- variables that cause spurious associations between treatments (user profile attributes) and outcome (user susceptibility). When left out, confounders can result in biased and inconsistent effect estimations. 

But what is the main source of confounding bias in fake news dissemination? Various studies in psychology and social science have shown the strong relationships of user behavior with user characteristics and activities such as information sharing, personality traits and trust \cite{waheed2017investigation,buffardi2008narcissism,benevenuto2009characterizing}. Consequently, characterizing user behavior has become a vital means to analyzing activities on social networking sites. Informed by this, we argue that \textit{fake news sharing behavior}, i.e., the user-news dissemination relations characterized by a bipartite graph (see Figure \ref{framework} \textcircled{1}), is critical to address confounding in causal relations between user attributes and susceptibility.

Learning fake news sharing behaviour is challenging because virtually all observational social media data is subject to \textit{selection bias} due to self-selection (e.g., users typically follow what they like) and the actions of online news platforms (e.g., these platforms only recommend news that they believe to be of interest to the users) \cite{schnabel2016recommendations}. Consequently, these biased data only partially describe how users share fake news. To alleviate the selection bias, one can leverage a technique commonly used in causal inference \cite{imbens2015causal}, particularly, \textit{Inverse Propensity Scoring} (IPS) \cite{rosenbaum1983central} that creates a pseudo-population similar to data collected from an randomized experiment. In context of fake news, propensity describes the probability of a user being exposed to a fake news piece. By connecting fake news dissemination with causal inference, we can derive an unbiased estimator for learning fake news sharing behavior under selection biases. 

The main contribution of this work is three-fold. First, we address a novel and important problem that complements earlier efforts on fake news detection. In particular, we seek to answer \textit{why} people share fake news by uncovering the causal relationships between user profiles and susceptibility. Second, we show how learning fake news sharing behavior under selection biases can be approached with propensity-weighting techniques. We design three simple and effective estimations of propensity score for fake news dissemination -- News-, User-News- and Neural-Network-based -- to learn \textit{unbiased embeddings} of fake news sharing behavior. Third, under the \textit{multiple causal inference} framework with mild assumptions, we propose to use the learned embeddings of fake news sharing behavior as the confounder, drawing from findings in social science. This enables us to learn a causal model that can identify causal user attributes and estimate their effects on user susceptibility. 

Our contributions are validated in an extensive empirical evaluation\footnote{Code is available at \url{https://github.com/GitHubLuCheng/Causal-Understanding-of-Fake-News-Dissemination}.}. For the first task of modeling fake news dissemination, we show that our proposed unbiased estimators improve accuracy of predicting fake news that users are more likely to spread. By comparing the learned embeddings of fake and true news sharing behavior, we make insightful findings on the differences of the two sharing behaviors. For the second task of identifying causal attributes of susceptible users, we first show that the predictive accuracy can be improved by incorporating the unbiased embeddings of fake news sharing behavior as confounders. We then reveal multiple user attributes that are potential causes of user susceptibility. The study concludes with some critical theoretical and practical implications for researchers and policy makers.
\section{Related Work}
\subsection{Fake News Detection}
Established work generally falls in two categories: content-based and propagation-based methods \cite{zhou2018survey}. In content-based methods, news content is typically represented by knowledge, style, or a latent representation. Knowledge-guided methods seek to directly evaluate news authenticity by comparing its knowledge with that within a knowledge graph. Fake news detection then naturally becomes a link prediction task \cite{shi2016discriminative}. Limited to the completeness of knowledge graphs, further post-processing approaches for knowledge inference are often required \cite{nickel2015review}. Style features can be word-level features such as TF-IDF and/or LIWC features \cite{perez2017automatic,castelo2019topic}. However, style-based methods are ``rarely supported by fundamental theories across disciplines'' \cite{zhou2018survey}. Latent-representation-based methods (e.g., \cite{wang2018eann}) have limited interpretability. 

Propagation-based methods advocate the use of social context information. For instance, news cascade \cite{castillo2011information} was extended by introducing user roles (i.e., opinion leaders or normal users), stance (e.g., approval or doubt) and sentiments expressed in user posts \cite{wu2015false}. The underlying assumption is that the overall structure of fake news cascades differs from the true ones. In early detection of fake news, news cascade was used as multivariate time series to model the propagation path of each news story \cite{liu2018early}. Another line of research focuses on self-defined graphs such as a stance graph built on user posts \cite{jin2016news}. Fake news is then detected by mining the stance correlations within a graph optimization framework. A more common type of graphs explores relationship among news article, publishers, users, and user posts. For instance, PageRank-like algorithm \cite{gupta2012evaluating}, tensor and matrix factorization \cite{shu2019beyond}.

Despite the remarkable progress in detecting fake news, comparatively fewer efforts seek to understand what user profile attributes cause users to spread fake news. Here, we provide a novel \textit{causal understanding} by learning unbiased fake news sharing behavior. This study complements earlier works by explicitly modeling fake news dissemination with a focus on combating selection bias and discovering user attributes causally related to user susceptibility.
\subsection{Propensity Scoring Methods}
As one of the most important techniques in causal inference, propensity score has been applied to observational studies in various fields such as medicine, economics, and computer science. The goal of propensity scoring methods is to create a pseudo-randomized trial by reweighting samples in different treatment groups using propensity scores \cite{bonner2018causal} -- essentially a balancing score. One of the most classical propensity scoring methods is IPS \cite{rosenbaum1983central}, where a unit's weight is equal to the inverse of its propensity score. Among all applications, ones that are most relevant to our task are causal recommender system \cite{bonner2018causal,schnabel2016recommendations} and domain adaptation \cite{sugiyama2012machine,cheng2020representation}. Conventional recommender systems are subject to selection bias. Recent studies (e.g., \cite{schnabel2016recommendations,liang2016causal}) proposed to use IPS for unbiased evaluation and learning of recommender system. For instance, user preferences (inferred through ratings or user and item covariates) were used to learn unbiased estimators from biased rating data \cite{schnabel2016recommendations}.  

IPS has been similarly applied to domain adaptation and covariate shift. In particular, these methods reweighed the distributions of source and target domains to adjust for their distributional differences \cite{sugiyama2007covariate,huang2007correcting,cheng2020representation}. For instance, to address the sample selection bias, a nonparametric method was proposed to directly produce resampling weights without distribution estimation \cite{huang2007correcting}. Another interpretation of IPS is importance weighting (e.g., \cite{sugiyama2007covariate}). Under the covariate shift, standard model selection techniques do not work as desired. Methods such as importance weighted cross validation (IWCV) \cite{sugiyama2007covariate} employed IPS to alleviate misestimation due to covariate shift. More recent work (e.g., \cite{cheng2020representation}) further used IPS to learn domain invariant representations.  

Informed by successful prior studies, in this work, we propose to leverage IPS to learn unbiased fake news sharing behavior under selection biases. We further design three simple and effective formulations to estimate propensity score in fake news dissemination. In doing so, we seek to (1) identify the causal user attributes by conditioning on the learned fake news sharing behavior; (2) study the differences between the fake and true news sharing behavior; and (3) improve models' prediction accuracy.
\section{Problem Statement}
Let $\mathcal{U}=\{1,2,...,u,...,U\}$ denote users who share fake news $\mathcal{C}=\{1,2,...,i,...,N\}$. $Y_{ui}\in\mathcal{Y}$ is a binary variable representing interactions between user $u$ and fake news $i$: if $u$ spreads $i$, then $Y_{ui}=1$ else, $Y_{ui}=0$. Note that $Y_{ui}=0$ can be interpreted as either $u$ is not interested in $i$ or $u$ did not observe $i$. Suppose users have $m$ profile attributes denoted by matrix $\mat{A}=(\mat{A}_1,\mat{A}_2,...,\mat{A}_m)$. Each user $u$ is also associated with an outcome $B\in(0,1]$, denoting $u$'s susceptibility to spread fake news. We aim to identify causal user attributes and estimate the effects, which consist of two tasks:
\begin{itemize}[leftmargin=*]
\item \textbf{Fake News Sharing Behavior Learning}. \textit{Given the user group $\mathcal{U}$, the corpus of fake news $\mathcal{C}$, the set of user-fake news interactions $\mathcal{Y}$, we aim to model the fake news dissemination process and learn fake news sharing behavior $\mat{U}$} under selection biases;
\item \textbf{Causal User Attributes Identification}. \textit{Given the user attributes $\mat{A}$, the fake news sharing behavior $\mat{U}$, and the user susceptibility $B$, this task seeks to identify user attributes that potentially cause users to spread fake news and estimate the effects.}
\end{itemize}
\section{The Proposed Framework}
\label{sec framework}
\begin{figure*}
  \includegraphics[width=.8\linewidth]{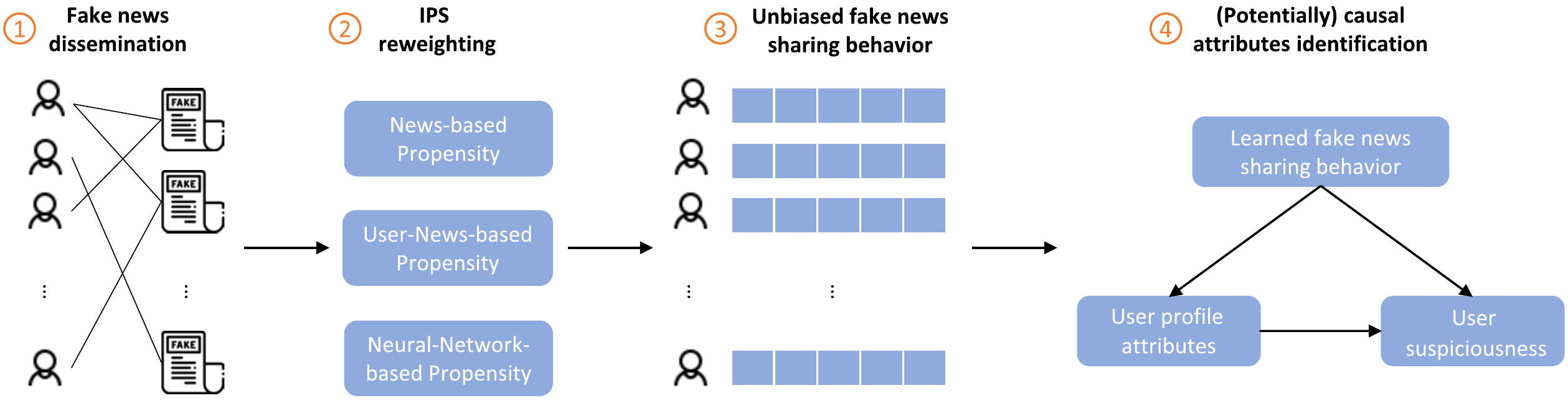}
    \caption{Overview of our framework. We model the fake news dissemination under selection biases (\textcircled{1}) and design three effective estimations of propensity score (\textcircled{2}) to learn unbiased embeddings of fake news sharing behavior (\textcircled{3}). Following the causal graph with the fake news sharing behavior being the confounder (\textcircled{4}), we examine the causal relationships between user profile attributes and susceptibility. Note that the identified attributes are ``potentially'' causal because as with most other observational studies, no \textit{conclusive} causal claims can be made.}
    \label{framework}
\end{figure*}

As with other observational studies, data for studying fake news is also subject to the common selection bias. In this section, we first provide mathematical formulations of the propensity-weighting model for fake news dissemination under selection biases. We then introduce three estimations of propensity score for learning unbiased embeddings of fake news sharing behavior. Under Potential Outcome framework \cite{rosenbaum1983central}, these embeddings are then used to identify the causal relationships between user attributes and susceptibility. Figure \ref{framework} features the overview of the proposed framework.
\subsection{Modeling Fake News Dissemination}
We begin by building a model that characterizes fake news dissemination. The key is the ``implicit'' feedback we collect through natural behavior such as news reading or sharing of a user with unique profile attributes.
By noting which fake news a user did and did not share in the past, we may infer fake news that a user will be interested in sharing in the future.
To better formulate the process of fake news dissemination, we introduce two binary variables highly related to this process: interestingness $R_{ui}\in\{0,1\}$ and exposure $O_{ui}\in\{0,1\}$. $R_{ui}=1(0)$ indicates $u$ is interested (not interested) in $i$; $O_{ui}=1$ denotes user $u$ was exposed to fake news $i$ and $O_{ui}=0$, otherwise. Therefore, we assume that a user spreads fake news iff s/he is both exposed to and interested in it \cite{joachims2017unbiased}:
\begin{equation}
Y_{ui}=O_{ui}\cdot R_{ui}, 
\end{equation}
\begin{equation}
\begin{split}
P(Y_{ui}=1)&=P(O_{ui}=1)\cdot P(R_{ui}=1),\\
&=\theta_{ui}\cdot\gamma_{ui} \quad \theta_{ui}>0; \gamma_{ui}>0;\forall Y_{ui} \in \mathcal{Y},
\end{split} 
\label{yui}
\end{equation}
where $\theta_{ui}=P(O_{ui}=1)$ and $\gamma_{ui}=P(R_{ui}=1)$ parameterize the probability of exposure and interestingness, respectively. As fake news dissemination is missing-not-at-random (MNAR)\footnote{MNAR implies the probability of an event (e.g., sharing fake news) being missing/unobserved varies for reasons that are unknown to us.} \cite{little2019statistical}, we further assume that the probability of $u$ spreading $i$ is represented as the product of the exposure and interestingness parameters \cite{saito2019unbiased}. 

Suppose we have a pair of fake news $(i,j)$ with $i\neq j$ and $\mathcal{D}_{pair}=\mathcal{U}\times\mathcal{C}\times\mathcal{C}$ is the set of all observed (positive) interactions $(u,i)$ and unobserved (negative) interactions $(u,j)$. As both the interestingness variable and exposure variable are \textbf{unobserved}, the model parameters are learned by optimizing the pairwise BPR (Bayesian Personalized Ranking) loss \cite{rendle2012bpr} that employs user-news interactions. In doing so, we assume that the observed user-news interactions better explain users' preferences than the unobserved ones, thereby, should be assigned higher prediction scores. We first define the ideal loss function of fake news dissemination as
\begin{equation}
    \mathcal{L}_{ideal}\big(\hat{\mat{S}}\big)=\frac{1}{|\mathcal{D}_{pair}|}\sum_{(u,i,j)\in \mathcal{D}_{pair}}\gamma_{ui}(1-\gamma_{uj})\ell\big(\hat{\mat{S}}_{uij}\big),
    \label{ideal}
\end{equation}
where $\hat{\mat{S}}_{uij}$ is the difference between the predicted scores of fake news $i$ and $j$, and $\ell=-\ln(\sigma(\cdot))$ represents the local loss for the triplet $(u,i,j)$. To this end, modeling fake news dissemination is a statistical estimation problem where we seek to estimate the ideal loss functions that returns news users are most interested in using the observed user-news interactions. 
\subsection{Learning Unbiased Sharing Behavior}
The previously introduced model for fake news dissemination directly employs user-news interactions collected from observational studies.
This leads to at least two major deficiencies: first, observational data only includes positive interactions between users and fake news whereas negative interactions are never observed. Consequently, the above fake news dissemination model cannot differentiate whether unshared fake news is uninteresting to the user or has yet to be exposed to the user; second, similar to the preferential attachment theory\footnote{Preferential attachment describes a phenomenon that the connection probability to an existing node is proportional to the degree of the target node.} \cite{barabasi1999emergence} in social network science, users are preferentially to interact with news that are already prevalent and online news platforms are also more likely to recommend popular news than the tail ones. Fake news dissemination models using these partially observed interactions will learn biased embeddings of the fake news sharing behavior (or user embeddings).

To handle selection bias, we propose to leverage IPS \cite{rosenbaum1983central,rubin2001using} to learn unbiased fake news sharing behavior based on existing positive interactions between users and fake news. To recall, propensity in fake news dissemination denotes the probability of exposing a user to a fake news piece. IPS works as a reweighting mechanism by assigning larger weights to news that is less likely to be observed.
Particularly, we assume that the event of user being exposed to fake news is probabilistic, i.e., the marginal probability $\theta_{ui}=P(O_{ui}=1)$ of observing a non-zero entry $Y_{ui}$ for all user-fake news pairs. Formally, we define the propensity score in the fake news dissemination as follows:
\begin{definition}[Propensity Score] The propensity score of user $u$ being exposed to news $i$ is
\begin{equation}
 \theta_{ui}=P(O_{ui}=1)=P(Y_{ui}=1|R_{ui}=1).
 \label{ps}
\end{equation}
\end{definition}
\noindent Eq. \ref{ps} indicates that the propensity score is the probability of $u$ spreading $i$ given $u$ is interested in $i$. This ensures that, in principle, there could be positive interaction between every pair of $(u,i)$. Incorporating $\theta_{ui}$ into the ideal loss function of fake news dissemination, we obtain the following unbiased estimator:
\begin{equation}
\hat{\mathcal{L}}_{unbiased}\big(\hat{\mat{S}}\big)=\frac{1}{|\mathcal{D}_{pair}|}\sum_{(u,i,j)\in \mathcal{D}_{pair}}\frac{Y_{ui}}{\theta_{ui}}\big(1-\frac{Y_{uj}}{\theta_{uj}}\big)\ell\big(\hat{\mat{S}}_{uij}\big),
\label{UES}
\end{equation}
Informed by the MNAR literature \cite{saito2019unbiased,joachims2017unbiased}, in the following proposition, we show that this propensity-score-based estimator is unbiased w.r.t. fake news dissemination.
\begin{proposition} The loss function in Eq. \ref{UES} is unbiased against the ideal loss of fake news dissemination in Eq. \ref{ideal}.
\begin{equation}
    \mathbb{E}\big[\hat{\mathcal{L}}_{unbiased}\big(\hat{\mat{S}}\big)\big]=\mathcal{L}_{ideal}\big(\hat{\mat{S}}\big).
\end{equation}
\end{proposition}
\noindent The proof of this proposition can be found in Appendix A. 
\subsubsection{Propensity Score for Fake News Dissemination} Here, we propose three estimations of propensity score based on user and news attributes.
The first formulation estimates propensity score using relative news popularity and is defined as 
\begin{definition}[News-based Propensity] Propensity using relative news popularity is defined as 
\begin{equation}
    P_{news}=\hat{\theta}^{news}_{,i}=\bigg(\frac{\sum_{u\in\mathcal{U}}Y_{ui}}{\max_{i\in\mathcal{C}}\sum_{u\in \mathcal{U}}Y_{ui}}\bigg)^{\eta},
    \label{p_news}
\end{equation}
\end{definition}
\noindent Typically, popularity-related measures follow power law distributions, therefore, we include the smoothing parameter $\eta\leq 1$ and set it to 0.5. 
With $P_{news}$, we assume that the probability of a user observing a fake news piece is highly related to its popularity.

\begin{definition}[User-News-based Propensity] Propensity using both relative news popularity and user popularity is defined as 
\begin{equation}
    P_{user}=\hat{\theta}^{user}_{u,i}=\bigg(\frac{\sum_{u\in\mathcal{U}}Y_{ui}\cdot F_u}{\max_{i\in\mathcal{C}}\sum_{u\in \mathcal{U}}Y_{ui}\cdot F_u}\bigg)^{\eta},
    \label{p_user}
\end{equation}
\end{definition}
\noindent where $F_u$ denotes the number of followers of $u$ and $\eta=0.5$.
$P_{user}$ also considers the bias induced by the user popularity, that is, users who are popular and active on social media are more likely to be exposed to fake news. Both estimations are input of Eq. \ref{UES}.
In the third formulation, we jointly estimate the propensity score and model fake news dissemination.
\begin{definition}[Neural-Network-based Propensity] Propensity encoded by neural networks is defined as 
\begin{equation}
    P_{neural}=\hat{\theta}^{neural}_{,i}= \sigma(\mat{e}_i),
    \label{p_neural}
\end{equation}
\end{definition}
\noindent where $\mat{e}_i$ is the latent representations of news content and $\sigma (\cdot)$ is the sigmoid function.
Here, we implicitly encode the popularity of fake news in the latent space based on the news content.
\subsubsection{Variance Reduction}
It is widely known that IPS-based approaches often suffer from large variance as the propensity score can be extremely small.
For example, fake news that is unpopular has low exposure probability. To reduce the variance, we employ the following non-negative loss \cite{saito2019unbiased}:
\begin{definition}[Non-Negative Loss] Given the propensity scores, the non-negative loss can be defined as 
\begin{equation}
\small
    \hat{\mathcal{L}}_{non-neg}\big(\hat{\mat{S}}\big)=\frac{1}{|\mathcal{D}_{pair}|}\sum_{(u,i,j)\in \mathcal{D}_{pair}}\max\bigg\{\ell_{unbiased}\big(\hat{\mat{S}}_{uij}\big),0\bigg\}.
    \label{nonneg}
\end{equation}
\end{definition}
\noindent Similar to the non-negative loss for Positive-Unlabeled learning with limited Positive data, Eq. \ref{nonneg} is more robust against the small propensity scores, and reduces the variance at the cost of introducing some bias \cite{kiryo2017positive}. The final loss function for modeling the unbiased fake news dissemination is formulated as follows:
\begin{equation}
    \arg\min_{\mat{U},\mat{V}}\hat{\mathcal{L}}_{non-neg}\big(\hat{\mat{S}}\big)+\lambda(\|\mat{U}\|^2_2+\|\mat{V}\|^2_2),
\end{equation}
\noindent where $\mat{U}$ and $\mat{V}$ are user embeddings (i.e., the embeddings of fake news sharing behavior) and news embeddings, respectively. $\lambda$ is a hyperparameter that controls the weight of the $\ell_2$-regularization for the latent factors.

\subsection{Identifying Causal User Attributes}
This section discusses how to simultaneously identify multiple user attributes that \textit{potentially} cause user susceptibility and estimate the effects. Causal inference is the anchor of knowledge to understand the underlying mechanism that drives people to spread fake news \cite{guo2020survey}. With multiple user attributes at hand, we are essentially tackling a multiple causal inference task where user attributes represent the multiple treatments and user susceptibility denotes the outcome. The goal is to estimate simultaneously the effects of individual user attributes on how likely a user spread a fake news piece. 

Suppose $u$'s attributes are encoded in a vector $\bm{a}=(a_1,a_2,...,a_m),$ $\bm{a}\in\mat{A}$. For each user $u$, there is a potential outcome function that maps configurations of the attributes to user susceptibility $B_u\in(0,1]$ which is formally defined as 
\begin{equation}
    B_u=n^u_{fake}\big/(n^u_{fake}+n^u_{true}),
\end{equation}
\noindent where $n^u_{fake}$ is the number of fake news $u$ has shared. Here we assume that \textit{a larger portion of news a user has shared is fake, more susceptible s/he is to share fake news}.

Multiple causal inference seeks to identify the sampling distribution of the potential outcomes $B_u(\bm{a})$ for each configuration of the attributes $\bm{a}$.
However, in observational studies, we can only observe one potential outcome of a user under one configuration of $\bm{a}$, a.k.a. the ``fundamental problem of causal inference'' \cite{holland1986statistics}. Without knowing the full distribution of $B_u(\bm{a})$ for any $\bm{a}$, the inference of the outcome can be biased, i.e., $\mathbb{E}[B_u(\bm{a})]\neq\mathbb{E}[B_u(\bm{a})|\mat{A}_u=\bm{a}].$ The key is to identify \textit{confounder} $\mat{Z}$ that simultaneously influences the causes $\mat{A}$ and the outcome $B$. We first introduce the following standard assumptions \cite{rosenbaum1983central,rubin1980randomization} in causal inference:
\begin{assumption}[Causal Inference Assumptions]\ 
\begin{enumerate}[leftmargin=*]
    \item The stable unit treatment value assumption (SUTVA): no interference between individuals and no different versions of a cause.
    \item The positivity or sufficient overlap assumption, that is 
    \begin{equation}
        0<p(\bm{a}_u\in\mathcal{A}|\mat{z}_u)<1,
    \end{equation}
    for all sets $\mathcal{A}$ with $p(\mathcal{A})>0$. It implies that given $\mat{Z}$, the conditional probability of any vector of the causes is positive.
    \item $\mat{Z}$ is sufficiently rich to capture all variables influencing both $\mat{A}$ and $B$:
    \begin{equation}
        p(\bm{a}_u\in\mathcal{A}|B_u(a_1),...,B_u(a_m),\mat{z}_u)=p(\bm{a}_u|\mat{z}_u).
    \end{equation}
\end{enumerate}
\end{assumption}
\noindent When applied to identifying user attributes that cause user susceptibility, two critical questions remain to be answered: 
\begin{enumerate}
    \item \textit{what are the variables causally related to both the user attributes and user susceptibility?} and
    \item \textit{In which cases, can all the three assumptions be satisfied?}
\end{enumerate}

For the first question, the key is to understand the positive interactions between users and fake news, i.e., the \textit{fake news sharing behavior} on social networking sites \cite{talwar2019people}. Decades of research in psychology and social science suggests that individual's online behavior and preferences are highly related to her personality traits \cite{buffardi2008narcissism}, cultural norms \cite{tsoi2011privacy}, and her social activities such as hate propagation \cite{rost2016digital}. Therefore, drawing from these findings, we propose to use the learned unbiased embeddings of fake news sharing behavior as the surrogate confounder, that is, $\mat{U}_u\approx\mat{Z}_u$. The underlying assumption is that users' behavior of sharing fake news is sufficient to explain both user attributes and user susceptibility. The corresponding causal graph is illustrated in \textcircled{4} in Figure \ref{framework}.

For the second question, to satisfy SUTVA, it is required that user susceptibility is independent of other users' attributes and same value of an attribute has the same interpretation for all users. For example, whether a user spreads fake news or not should not depend on the age and gender of any other user. The positivity assumption can be interpreted as the observed attributes values vary within the counfounder $\mat{Z}$ strata, i.e., there should be adequate exposure variability of different levels of user attributes within the strata of fake news sharing behavior. The third assumption requires the learned embedding of the fake news sharing behavior to account for all confounding bias. 

Given user profile attributes $\mat{A}$, confounder $\mat{U}$ and the user susceptibility $B$, we build the causal model to identify the causal user attributes and estimate their effects:
\begin{equation}
    B_u=\bm{\beta}^\intercal\bm{a}_u+\bm{\gamma}^\intercal\mat{U}_u,
    \label{outcome}
\end{equation}
where $\bm{\beta}$ and $\bm{\gamma}$ are coefficients. $\bm{\beta}$ denotes how user attributes affect individual decision to share fake news. A positive coefficient in $\bm{\beta}$ that passes statistical significance test indicates users with larger value on this attribute are more susceptible to spread fake news.  
\section{Empirical Evaluation}
The empirical evaluation starts with descriptions of experimental setup, including the datasets, baselines, evaluation metrics, and implementation details. In the second part, we report results of our experiments and discuss the implications. 
\begin{table}
\begin{center}
\caption{Dataset statistics. }
\begin{tabular}{ c|c|c|c|c } \hline
Dataset & \# Real &\# Fake &\# Total&\# Users \\ \hline
\textit{PolitiFact}&624 &432 & 1,056& 110,127 \\ \hline
\textit{GossipCop}&16,817&5,323&22,140&194,788\\ \hline
\end{tabular}
\label{data}
\end{center}
\end{table}
\subsection{Experimental Setup}
\textbf{Data.} Two benchmark datasets\footnote{Both are available at https://github.com/KaiDMML/FakeNewsNet.} for fake news detection are used for evaluation: PolitiFact\footnote{https://www.politifact.com/} and GossipCop\footnote{https://www.gossipcop.com/}.  
\begin{itemize}[leftmargin=*]
    \item \textit{PolitiFact.} In \textit{PolitiFact}, political news was collected from various sources and fact-checking evaluation results, i.e., fake or real, are provided by journalists and domain experts. This dataset consists of 624 real news and 432 fake news.
    \item \textit{GossipCop.} In \textit{GossipCop}, entertainment stories were collected from various media outlets. The fact-checking evaluation results came from the rating scores on the GossipCop website. Ratings range from 0 to 10 with 0 indicating fake and 10 real. Different from PolitiFact, GossipCop intends to show more fake stories due to its entertainment purpose. The dataset consists of 16,817 real stories and 5,323 fake stories.
\end{itemize}
\noindent The basic statistics of these two datasets are shown in Table \ref{data}. For each dataset, we create the training and test datasets with a 80/20 split. The training data is randomly selected from the original data (thus biased) whilst from the rest data, we create the test data such that we expose each user to each fake news as uniformly as possible (i.e., with equal probability, thus less biased). This method, which has been advocated as the most practical way to imitate randomized experiments \cite{bonner2018causal,liang2016causal}, can generate data with users' decisions under random exposures.
The evaluation of causal models has long been a challenging task due to the lack of ground truth.
By creating the distributional differences between the training and test data, we can compare a causal method with a non-causal method using the prediction accuracy across different environments.
A causal method is expected to be more robust to the distribution shift as it is more transportable and domain-invariant \cite{pearl2011transportability,peters2015causal}.   

\noindent\textbf{Baselines.} We are not aware of any similar work in the literature of fake news that learns the embeddings of fake news sharing behavior and identifies causal user profile attributes.
As our problem setting is closely related to recommender systems, here, we employ two standard approaches in recommender systems with implicit feedback as backbones of our model: Bayesian personalized ranking for matrix factorization (BPRMF) \cite{rendle2012bpr} and the neural collaborative filtering model (NCF) \cite{he2017neural}. Note that for each baseline, our approach has three different variants corresponding to the three estimated propensity scores. For example, we incorporate the propensity scores defined in Eq. \ref{p_news}-\ref{p_neural} into BPRMF and get three different variants of our model: BPRMF-N, BPRMF-U and BPRMF-Neu.

We adopt two standard evaluation metrics in recommender systems -- Recall@K and NDCG@K. Recall@K measures of all fake news that were actually interesting to a user, how many the model predicted to be interesting in the top $K$ fake news. It focuses on the ratio of interesting fake news that are not missed by the algorithm.
NDCG@K measures the accuracy of the algorithm based on the ground truth interestingness and the predicted ranking of fake news among those ranked as the top $K$ interesting fake news. Their formal definitions can be found in Appendix B.
For the implementation, we used Tensorflow \cite{abadi2016tensorflow} and Statsmodel \cite{seabold2010statsmodels}. $\lambda$ is set to 1e-2 for BPRMF-based models and 1e-3 for NCF-based models, the embedding dimension is 64 and the batch size is 1,024 for both. We employ the plain architecture of NCF, where the dimension of each hidden layer keeps the same. All the models are optimized by RMSProp Optimizer \cite{tieleman2012lecture} with a learning rate of 1e-3 for BPRMF-based models and 1e-2 for NCF-based models. More implementation details can be found in Appendix B. 
\subsection{Evaluation on Fake News Dissemination}
We first evaluate models for learning fake news sharing behavior.
Specifically, we aim to answer the research questions below: 
\begin{itemize}[leftmargin=*]
    \item How does the proposed model fare against standard recommendation models w.r.t. the performance of predicting fake news that users will share?
    \item How is the fake news sharing behavior different from the true news sharing behavior in the latent space?
\end{itemize}
\noindent With the distribution shift between the training and test data, the first question examines the efficacy of the proposed IPS-reweighting models. For the second question, we first visualize the learned embeddings of the fake and true news sharing behavior in the 2-D space. We then compute the Silhouette Coefficient of the clustering results based on these embeddings.
\subsubsection{How does our approach fare against baselines in predicting fake news that users will spread?}
We compare Recall@K and NDCG@K of the two base models (i.e., BPRMF and NCF) to our models using the training and test data with distributional differences. We present the results averaged over 5 repetitions along with the relative improvement for both datasets in Table \ref{b_politic}-\ref{n_gossip}. The presented improvement on each dataset and for each evaluation measure is significant at 0.05 level. We begin by observing that indeed the imposed IPS reweighting confers an advantage to alleviating the selection bias in fake news dissemination, see, e.g., the results for \textit{PolitiFact} with the base model NCF in Table \ref{n_politic}. The improvement is most significant when $K$ is small, e.g., $K=20$. This indicates that our IPS-reweighting strategy is more effective when predicting fake news that is highly likely to be shared. All three IPS estimators can achieve the best performance w.r.t. Recall@K and NDCG@K with no evidence showing that one is most superior. User-News- and Neural-Network-based propensity, mostly, present better performance when predicting fake news across different environments. Estimating propensity using user popularity and news content may be more effective than using news popularity alone. 
\begin{table}
\caption{Performance comparisons w.r.t. predicting fake news to be shared using data \textit{PolitiFact} and base model BPRMF (\%). $p<0.05.$}
\begin{subtable}{\columnwidth}
\caption{Recall@K with K=20,40,60,80.}
\resizebox{\columnwidth}{!}{
\begin{tabular}{|l|l|l|l|l|}
\hline
K         & 20         & 40         & 60         & 80\\ \hline
BPRMF     & 12.36 & 22.18 & 31.10& 39.51\\ \hline\hline
BPRMF-N   & 14.45$^{\uparrow 16.9\%}$& 25.11$^{\uparrow 13.2\%}$& 34.34$^{\uparrow 10.4\%}$ & 42.72$^{\uparrow 8.1\%}$ \\ \hline
BPRMF-U   & 14.78$^{\uparrow 19.6\%}$ & 25.65$^{\uparrow 15.6\%}$ & 34.91$^{\uparrow 12.2\%}$ & \textbf{43.63$^{\uparrow 10.4\%}$}\\ \hline
BPRMF-Neu & \textbf{14.90$^{\uparrow 20.6\%}$} & \textbf{25.83$^{\uparrow 16.5\%}$} &\textbf{35.13$^{\uparrow 13.0\%}$} &43.55$^{\uparrow 10.2\%}$\\ \hline
\end{tabular} }
\end{subtable}
\begin{subtable}{\columnwidth}
\caption{NDCG@K with K=20,40,60,80.}
\resizebox{\columnwidth}{!}{
\begin{tabular}{|l|l|l|l|l|}
\hline
K         & 20        & 40        & 60         & 80   \\ \hline
BPRMF     & 5.33 & 7.51 & 9.22  & 10.71  \\ \hline\hline
BPRMF-N   & 6.39$^{\uparrow 19.9\%}$ & 8.73$^{\uparrow 16.2\%}$& 10.49$^{\uparrow 13.8\%}$ & 11.97$^{\uparrow 11.8\%}$ \\ \hline
BPRMF-U   & \textbf{6.54}$^{\uparrow 22.7\%}$ & 8.92$^{\uparrow 18.8\%}$ &10.69$^{\uparrow 15.9\%}$ & \textbf{12.21}$^{\uparrow 14.0\%}$\\ \hline
BPRMF-Neu & 6.53$^{\uparrow 22.5\%}$& \textbf{8.93}$^{\uparrow 18.9\%}$ & \textbf{10.71}$^{\uparrow 16.2\%}$ & 12.19$^{\uparrow 13.8\%}$\\ \hline
\end{tabular} }
\end{subtable}
\label{b_politic}
\end{table}
\begin{table}
\caption{Performance comparisons w.r.t. predicting fake news to be shared using data \textit{PolitiFact} and base model NCF (\%). $p<0.05.$}
\begin{subtable}{\columnwidth}
\caption{Recall@K with K=20,40,60,80.}
\resizebox{\columnwidth}{!}{
\begin{tabular}{|l|l|l|l|l|}
\hline
K         & 20         & 40         & 60         & 80       \\ \hline
NCF     & 9.59  & 18.45 & 27.30 & 36.33\\ \hline\hline
NCF-N   & \textbf{10.42}$^{\uparrow 8.7\%}$ & \textbf{19.34}$^{\uparrow 4.8\%}$& 28.58$^{\uparrow 4.7\%}$& 37.07$^{\uparrow 2.0\%}$ \\ \hline
NCF-U   & 10.29$^{\uparrow 7.3\%}$& 19.29$^{\uparrow 4.6\%}$ & 27.34$^{\uparrow 0.1\%}$& 34.87$^{\downarrow 4.0\%}$ \\ \hline
NCF-Neu & 10.20$^{\uparrow 6.4\%}$& 19.11$^{\uparrow 3.6\%}$ & \textbf{28.74}$^{\uparrow 5.3\%}$ & \textbf{38.39}$^{\uparrow 5.7\%}$\\ \hline
\end{tabular}}
\end{subtable}
\begin{subtable}{\columnwidth}
\caption{NDCG@K with K=20,40,60,80.}
\resizebox{\columnwidth}{!}{
\begin{tabular}{|l|l|l|l|l|}
\hline
K         & 20        & 40        & 60        & 80\\ \hline
NCF     & 3.72 & 5.66 & 7.35 & 8.94\\ \hline\hline
NCF-N   & 4.13$^{\uparrow 11.2\%}$& 6.09$^{\uparrow 7.6\%}$& \textbf{7.85}$^{\uparrow 6.8\%}$& 9.36$^{\uparrow 4.7\%}$\\ \hline
NCF-U   & \textbf{4.19}$^{\uparrow 12.6\%}$& \textbf{6.18}$^{\uparrow 9.2\%}$& 7.75$^{\uparrow 5.4\%}$ & 9.10$^{\uparrow 1.8\%}$\\ \hline
NCF-Neu & 4.04$^{\uparrow 8.6\%}$ & 5.99$^{\uparrow 5.8\%}$ & 7.82$^{\uparrow 6.4\%}$& \textbf{9.52}$^{\uparrow 6.5\%}$ \\ \hline
\end{tabular}}
\end{subtable}
\label{n_politic}
\end{table}
\subsubsection{Comparing News Sharing Behavior} 
\begin{figure}
\centering
\begin{subfigure}{.5\columnwidth}
\centering
\captionsetup{justification=centering}
  \includegraphics[width=.8\linewidth]{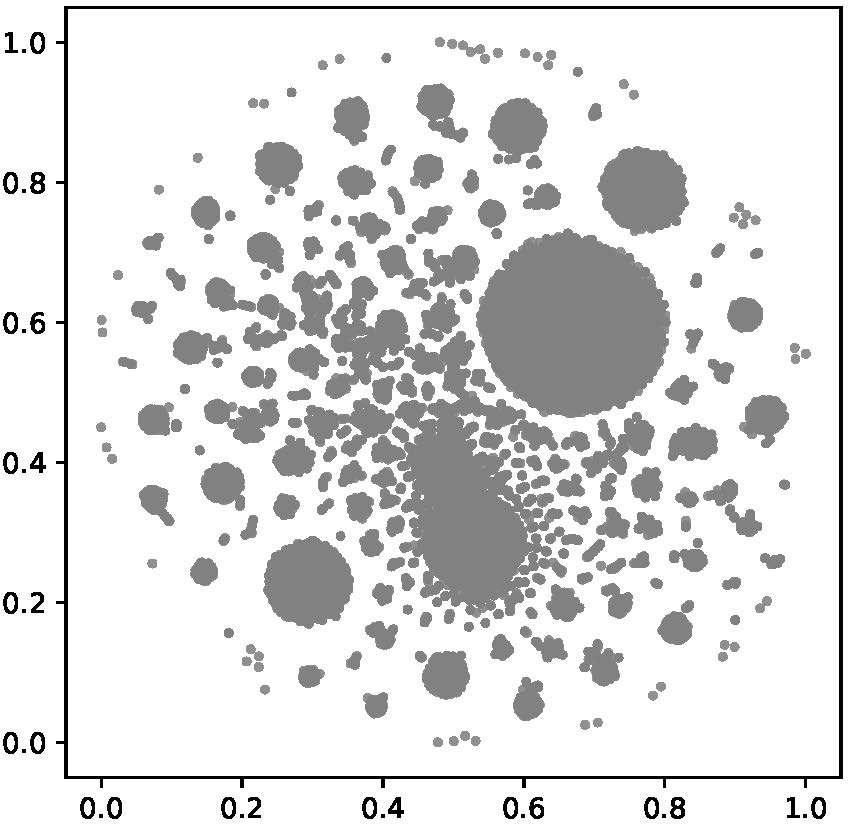}
  \caption{Fake news sharing behavior. Silhouette Coefficient=-0.124}
\end{subfigure}%
\begin{subfigure}{.5\columnwidth}
\centering
\captionsetup{justification=centering}
  \includegraphics[width=.8\linewidth]{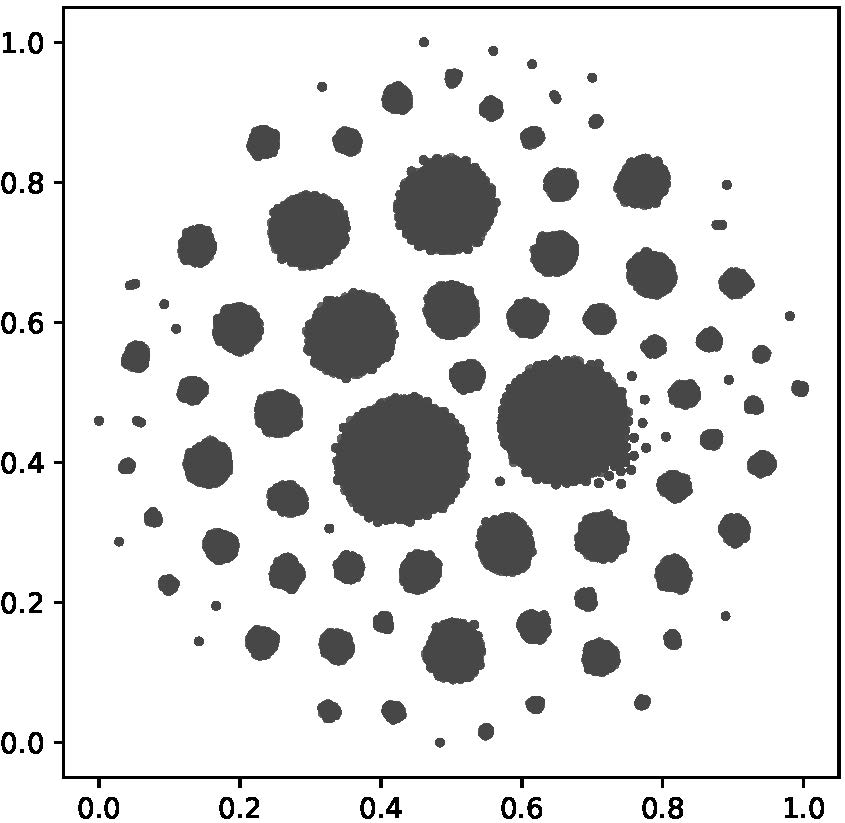}
    \caption{True news sharing behavior. Silhouette Coefficient=0.903}
\end{subfigure}
\caption{Behavior comparisons using 2-D t-SNE visualizations.}
\label{tsne_p}
\end{figure}

To learn the unbiased user behavior of sharing true news, we apply the same news dissemination model described in Section \ref{sec framework} to all true news and the associated users. We then extract embeddings of users who shared fake news and who only shared true news, and denote them as $\mat{U}_f$ and $\mat{U}_t$, respectively. We run BPRMF-N on \textit{PolitiFact} as a working example and visualize $\mat{U}_f$ and $\mat{U}_t$ in 2-D space using t-SNE \cite{maaten2008visualizing}. To ensure fair comparisons, we select users who only spread fake/true news and further conduct random sampling to make the number of both types of users equal (49,000 users). In addition to qualitative analysis, we further performed DBSCAN \cite{ester1996density} clustering on $\mat{U}_f$ and $\mat{U}_t$, respectively. Then we compute the Silhouette Coefficient of the inferred clusters. Results are presented in Figure \ref{tsne_p}.

An important notion is that embeddings of fake news sharing behavior are more concentrated on a single primary cluster whilst those of true news sharing behavior are better separated into multiple and smaller clusters. This is also evidenced by the results of Silhouette Coefficient, value of which ranges from -1 to 1. A larger value denotes that a sample is further away from its neighboring clusters. The Silhouette Coefficient of true news sharing behavior is close to $1$, indicating that the samples are well matched to their own clusters. We conclude that fake and true news sharing behavior are essentially different, also suggested by previous findings about fake news cascade \cite{castillo2011information}. Particularly, users who spread true news present more diverse behaviors whereas those spreading fake news have similar sharing behaviors. Our conclusion also echoes recent findings in social science and psychology \cite{talwar2019people} showing that people susceptible to spread fake news share key characteristics such as self-disclosure \cite{eder1991structure}) and social comparison \cite{keefer1994portrait}.
\begin{table}
\caption{Performance comparisons w.r.t. predicting fake news to be shared using data \textit{GossipCop} and base model BPRMF (\%). $p<0.05.$}
\begin{subtable}{\columnwidth}
\caption{Recall@K with K=20,40,60,80.}
\resizebox{\columnwidth}{!}{
\begin{tabular}{|l|l|l|l|l|}
\hline
K         & 20         & 40         & 60         & 80    \\ \hline
BPRMF     & 13.31 & 16.38 & 18.77& 20.8 \\ \hline\hline
BPRMF-N   & 14.92$^{\uparrow 12.2\%}$ & 17.61$^{\uparrow 7.5\%}$ & 19.70$^{\uparrow 5.0\%}$ & 21.52$^{\uparrow 3.5\%}$\\ \hline
BPRMF-U   & 14.97$^{\uparrow 12.6\%}$ & 17.70$^{\uparrow 8.1\%}$& 19.73$^{\uparrow 5.1\%}$& 21.58$^{\uparrow 3.8\%}$\\ \hline
BPRMF-Neu &\textbf{15.72}$^{\uparrow 18.2\%}$ & \textbf{18.76}$^{\uparrow 14.5\%}$ & \textbf{21.03}$^{\uparrow 12.0\%}$ & \textbf{22.96}$^{\uparrow 10.4\%}$\\ \hline
\end{tabular} }
\end{subtable}
\begin{subtable}{\columnwidth}
\caption{NDCG@K with K=20,40,60,80.}
\resizebox{\columnwidth}{!}{
\begin{tabular}{|l|l|l|l|l|}
\hline
K         & 20         & 40         & 60          & 80  \\ \hline
BPRMF     & 10.52 & 11.32& 11.86 & 12.30\\ \hline\hline
BPRMF-N   & 12.38$^{\uparrow 17.7\%}$& 13.11$^{\uparrow 15.8\%}$ & 13.60$^{\uparrow 14.7\%}$  & 13.97$^{\uparrow 13.6\%}$\\ \hline
BPRMF-U   & 12.22$^{\uparrow 16.2\%}$ & 12.95$^{\uparrow 14.4\%}$& 13.42$^{\uparrow 13.2\%}$ & 13.81$^{\uparrow 12.3\%}$ \\ \hline
BPRMF-Neu & \textbf{12.74}$^{\uparrow 21.1\%}$ &\textbf{13.56}$^{\uparrow 19.8\%}$ & \textbf{14.08}$^{\uparrow 18.7\%}$ & \textbf{14.49}$^{\uparrow 17.8\%}$ \\ \hline
\end{tabular} }
\end{subtable}
\label{b_gossip}
\end{table}
\begin{table}
\caption{Performance comparisons w.r.t. predicting fake news to be shared using data \textit{GossipCop} and base model NCF (\%). $p<0.05.$}
\begin{subtable}{\columnwidth}
\caption{Recall@K with K=20,40,60,80.}
\resizebox{\columnwidth}{!}{
\begin{tabular}{|l|l|l|l|l|}
\hline
K         & 20        & 40         & 60         & 80   \\ \hline
NCF     & 5.87 & 8.01 & 9.72  & 11.63 \\ \hline\hline
NCF-N   & 7.59$^{\uparrow 29.3\%}$& 9.50$^{\uparrow 18.6\%}$  & 11.22$^{\uparrow 15.4\%}$ & 12.74$^{\uparrow 9.5\%}$\\ \hline
NCF-U   & \textbf{8.99}$^{\uparrow 53.2\%}$ & \textbf{10.93}$^{\uparrow 36.5\%}$ & \textbf{12.73}$^{\uparrow 31.0\%}$ & \textbf{14.42}$^{\uparrow 24.0\%}$\\ \hline
NCF-Neu & 8.36$^{\uparrow 42.4\%}$ & 10.53$^{\uparrow 31.5\%}$ & 12.39$^{\uparrow 27.5\%}$ & 13.97$^{\uparrow 20.1\%}$\\\hline
\end{tabular} }
\end{subtable}
\begin{subtable}{\columnwidth}
\caption{NDCG@K with K=20,40,60,80.}
\resizebox{\columnwidth}{!}{
\begin{tabular}{|l|l|l|l|l|}
\hline
K         & 20        & 40        & 60        & 80     \\ \hline
NCF     & 4.41& 4.97 & 5.37& 5.77 \\ \hline\hline
NCF-N   & 5.96$^{\uparrow 35.1\%}$ & 6.50$^{\uparrow 30.8\%}$& 6.91$^{\uparrow 28.7\%}$ & 7.23$^{\uparrow 25.3\%}$\\\hline
NCF-U   & \textbf{7.36}$^{\uparrow 66.9\%}$ &\textbf{7.91}$^{\uparrow 59.2\%}$ & \textbf{8.33}$^{\uparrow 55.1\%}$& \textbf{8.68}$^{\uparrow 50.4\%}$\\ \hline
NCF-Neu & 6.53$^{\uparrow 48.1\%}$ & 7.14$^{\uparrow 43.7\%}$ & 7.57$^{\uparrow 41.0\%}$ & 7.91$^{\uparrow 37.1\%}$ \\ \hline
\end{tabular}}
\end{subtable}
\label{n_gossip}
\end{table}
\subsection{Evaluation on Identifying Causal User Attributes}
We show empirical results for identifying user profile attributes that potentially cause user susceptibility to share fake news.
With the unbiased embeddings of fake news sharing behavior as the confounder, in this experiment, we seek to (1) assess the effectiveness of outcome model Eq. \ref{outcome} by predicting user susceptibility; meanwhile (2) discover the causal user attributes and estimate the effects.
We thereby feed $B_u$ along with $\bm{a}_u$ and $\mat{U}_u$ into Eq. \ref{outcome}. All the experiments in this subsection are based on the BPRMF model. 
\subsubsection{Effect on Predicting User Susceptibility} 
\begin{figure}
\centering
\begin{subfigure}{.5\columnwidth}
\centering
\captionsetup{justification=centering}
  \includegraphics[width=.9\linewidth]{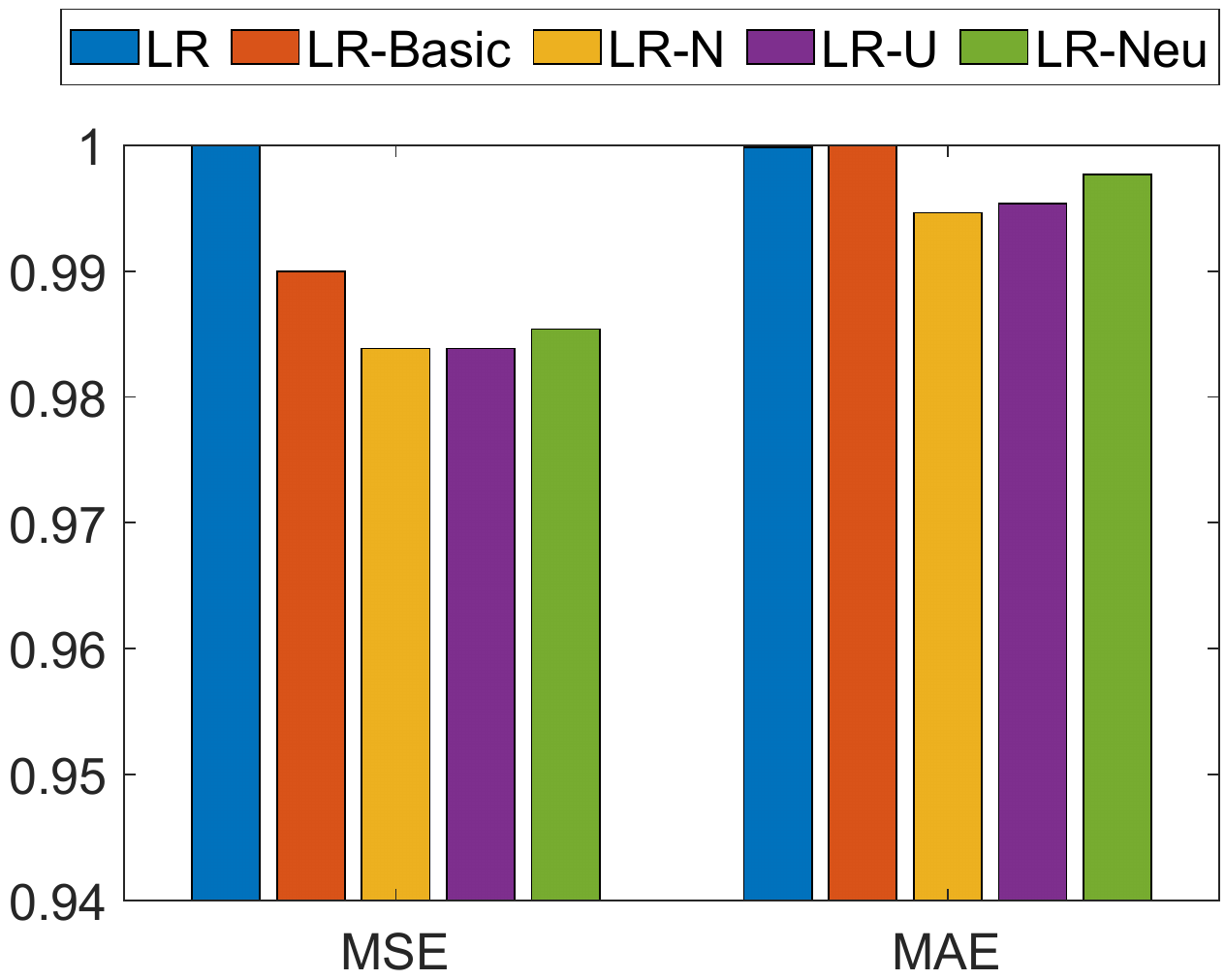}
  \caption{\textit{PolitiFact}.}
\end{subfigure}%
\begin{subfigure}{.5\columnwidth}
\centering
\captionsetup{justification=centering}
  \includegraphics[width=.9\linewidth]{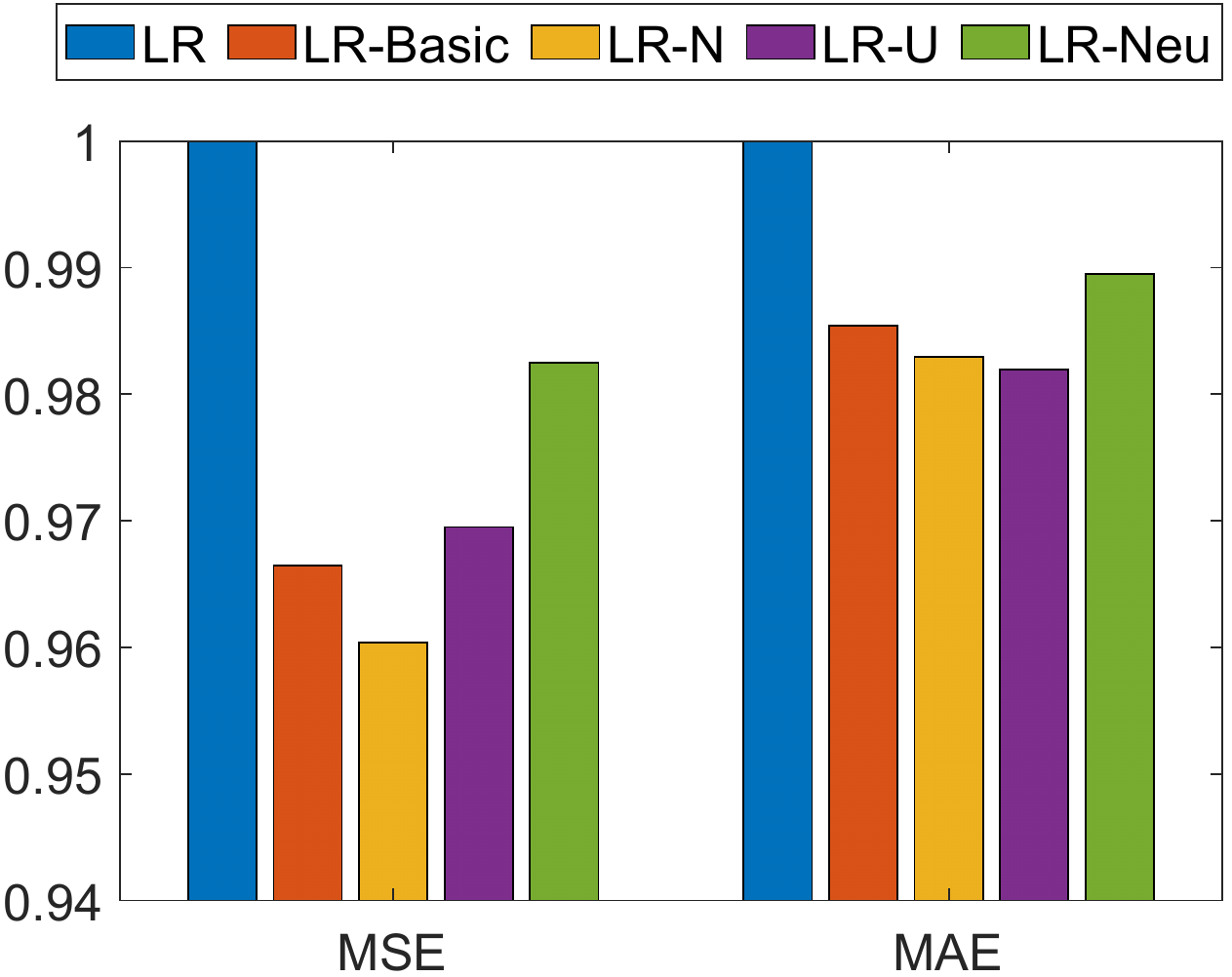}
    \caption{\textit{GossipCop}.}
\end{subfigure}
\caption{Performance comparisons w.r.t. predicting user susceptibility using both datasets. $y-$axis denotes relative results.}
\label{pred_user}
\end{figure}

As the focus of this experiment is to testify the effectiveness of the unbiased fake news sharing behavior on improving predictive accuracy, here, we take the simple Linear Regression (LR)\footnote{This experiment can be easily adapted to other machine learning models.} as the basic model and compare the performance of LR with various input: 
\begin{itemize}[leftmargin=*]
    \item \textit{LR}. The input solely consists of the user attributes.
    \item \textit{LR-Basic}. The input includes both the user attributes and embeddings of user sharing behavior learned via BPRMF.
    \item \textit{LR-N}. The input includes both the user attributes and embeddings of user sharing behavior learned via BPRMF-N.
    \item \textit{LR-U}. The input includes both the user attributes and embeddings of user sharing behavior learned via BPRMF-U.
    \item \textit{LR-Neu}. The input includes both the user attributes and embeddings of user sharing behavior learned via BPRMF-Neu.
\end{itemize}
We create the training and test data with a 80/20 split: 80\% users are in the training dataset. We report the two widely used evaluation metrics for regression -- Mean Squared Error (MSE) and Mean Absolute Error (MAE). The relative results are presented in Figure \ref{pred_user}. We begin by observing that the learned embeddings of user sharing behavior can improve the accuracy of predicting user susceptibility, see, e.g., the results for \textit{GossipCop}. Further, when taking the input of unbiased embeddings, LR can achieve the best results, especially for LR-N and LR-U. We may conclude that when predicting user susceptibility, incorporating the unbiased embeddings of fake news sharing behavior as the confounder has more positive influence on standard predictive models compared to biased embeddings.  
\subsubsection{Identification of Causal User Profile Attributes} 
\begin{figure}
\centering
  \includegraphics[width=.9\linewidth]{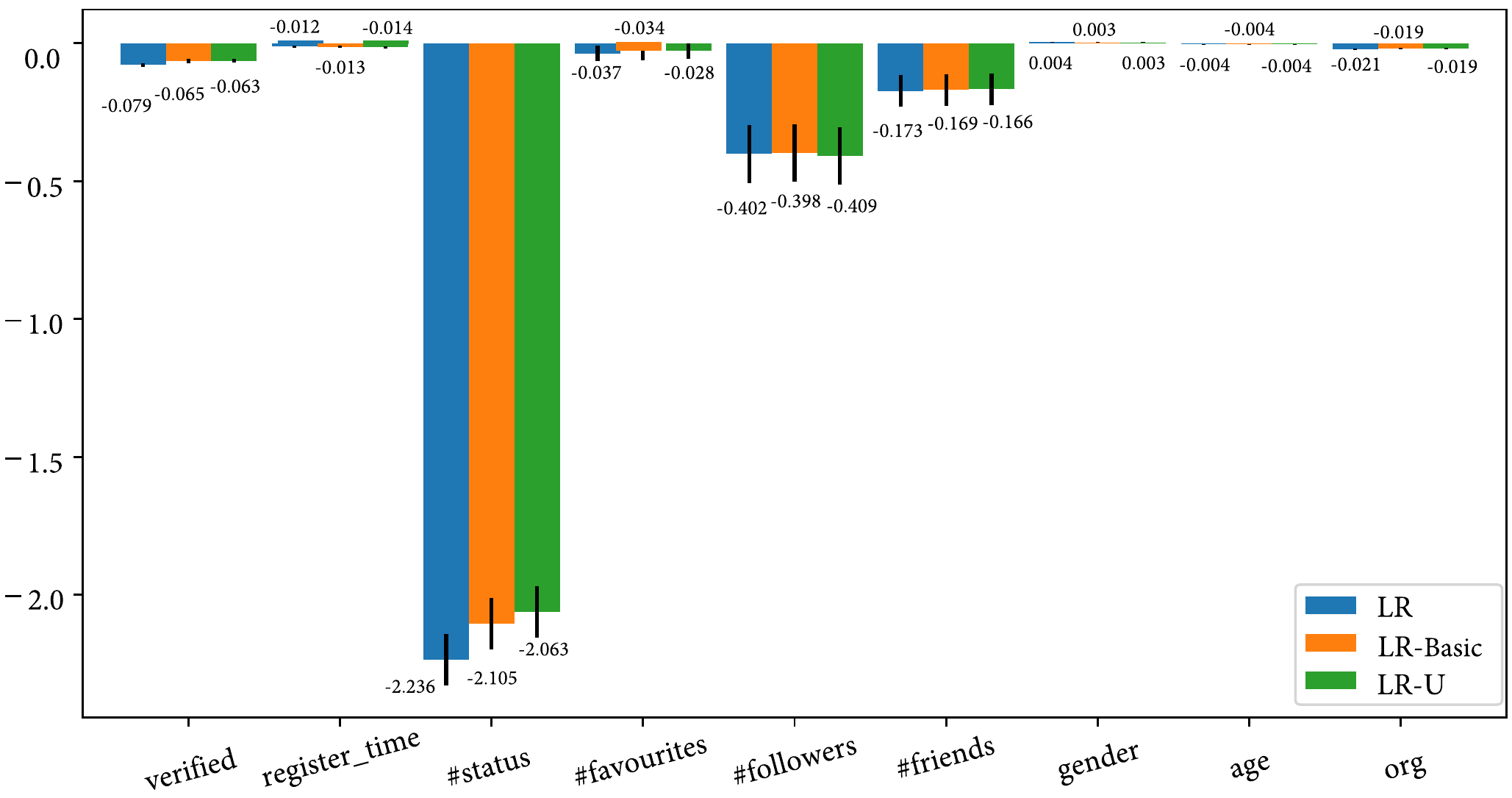}
    \caption{\textit{PolitiFact:} Effects comparisons w.r.t. each potential causal user attribute. All the results are statistically significant.}
    \label{cause_p}
\end{figure}

\begin{figure}
  \includegraphics[width=.9\linewidth]{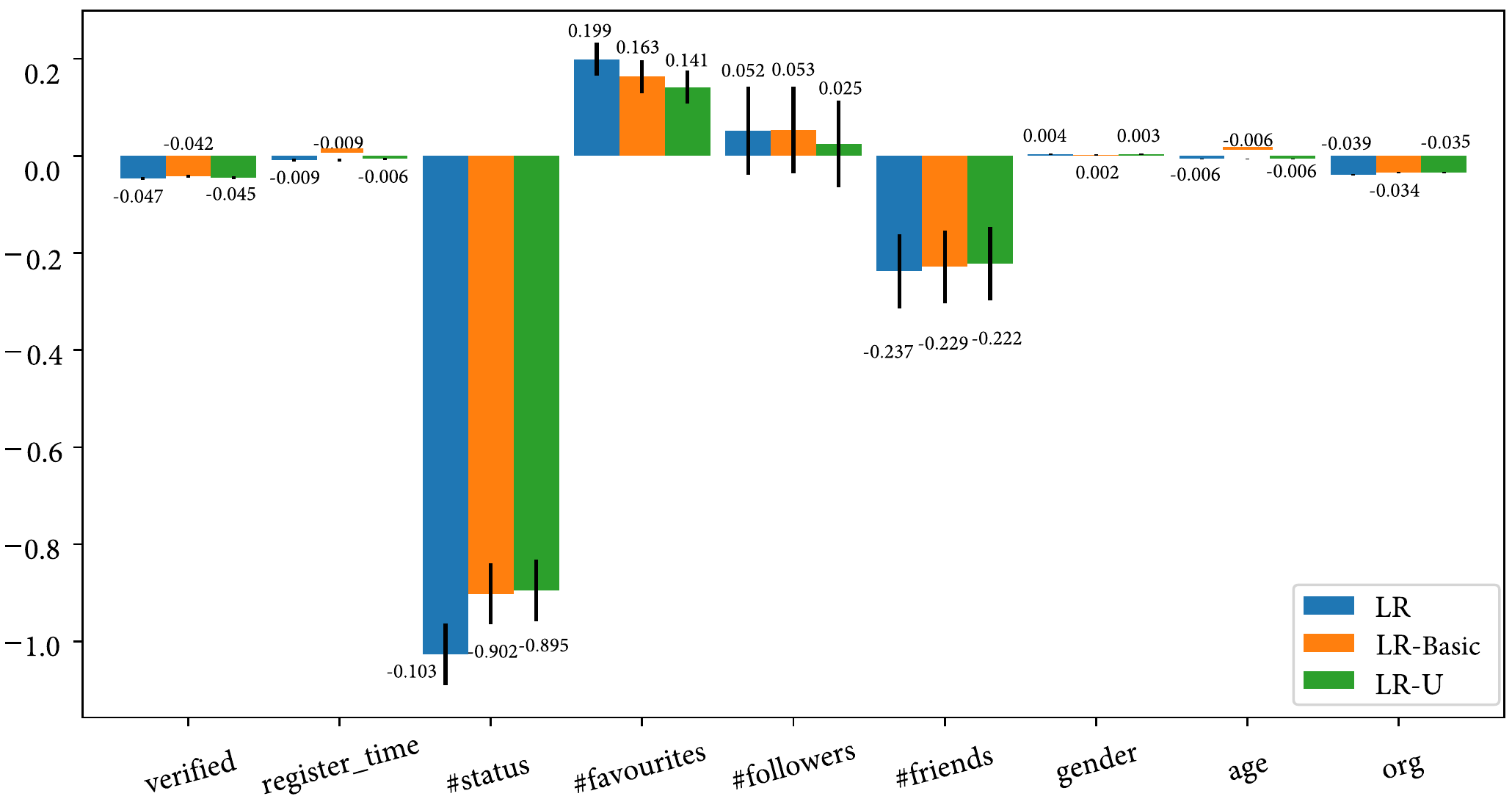}
    \caption{\textit{GossipCop:} Effects comparisons w.r.t. each potential causal user attribute. All the results except for that of \textit{\#followers} are statistically significant.}
    \label{cause_g}
\end{figure}

In this experiment, we examine the causal relationships between user attributes and user susceptibility and estimate the effects. Specifically, we compare the coefficients of \textit{LR-U} (i.e., debiased causal model) with those of \textit{LR-Basic} (i.e., biased causal model) and \textit{LR} (i.e., noncausal model). We use \textit{LR-U} as an example because similar comparison results can be found using \textit{LR-N} and \textit{LR-Neu}. We present the coefficients and confidence intervals in Figures \ref{cause_p}-\ref{cause_g}.

We first observe that \textit{LR-U} presents more conservative estimations of the effects, see, e.g., \textit{\#status} -- the number of Tweets (including retweets)
issued by the user -- for both datasets. This is partly because the unbiased embeddings can better alleviate the influence of confounding bias on the outcome. Additionally, \textit{\#status} has the largest effect on identifying a susceptible user, and the causal effect is negative. We may infer that users who have historically issued more tweets (regardless they are fake or not) are less susceptible to spread fake news. Similar negative effect can be observed in the binary attribute \textit{verified} (1 denotes verified) -- whether the user has a verified account, \textit{org} (1 denotes the account represents an organization) -- whether the account belongs to an organization, and \textit{\#friends} -- the number of users this account is following. Intuitively, verified users and organizations are less susceptible to share fake news. While lacking ground truth for causal user attributes, by identifying profile attributes that are intuitively causes, our causal models might be applied to discovering more intrinsic user attributes that describe why people share fake news.

Our results also align well with previous findings in psychology that users with more friends share less fake news because they seek to build positive image when comparing with peers \cite{talwar2019people}. Of particular interest is that there are two contradictory results across the two datasets: effects of both \textit{\#favorites} -- the number of Tweets users have liked -- and \textit{\#followers} are negative in \textit{PolitiFact} but become positive in \textit{GossipCop}. We surmise that (1) based on the causal transpotability theory, these two attributes are less likely to be the causes of user susceptibility to share fake news. Typically, we do not need a user's approval to follow him/her on social media and following is a one-way street; (2) the category of online news platforms is a possible confounder that is left out by the surrogate confounder. While significant causal relationships are found w.r.t. \textit{gender}, \textit{age}, and \textit{register\_time}, the nearly zero effects warrant further studies to make causal claims regarding these attributes. This is also supported by previously conducted survey studies such as \cite{buchanan2020people}.  

To summarize, the empirical evaluation shows that our proposed framework can learn unbiased embeddings of fake news sharing behavior that lead to more accurate predictions of fake news that users will share and user susceptibility. Our proposed causal framework also enables us to identify user attributes that potentially cause user susceptibility and estimate their effects. Comparisons between unbiased embeddings of true and fake news sharing behavior yield interesting findings regarding the differences between the two types of user behaviors. 
\section{Discussion}
We discuss the importance of understanding the causal relationships between user profile attributes and user susceptibility in combating the growing concerns about fake news. The results shown in this work demonstrate the efficacy of IPS-weighted news sharing models for learning unbiased fake news sharing behavior and the causal regression models for identifying user attributes potentially causing user susceptibility. While social media data, by itself, is not able to reliably identify the causes for why people share fake news, it can provide supporting evidence for existing conjectures and generate hypotheses for further investigation. 

The observation that IPS-weighted models consistently outperform the biased fake news dissemination models in predicting fake news a user is likely to spread suggests that causal inference theories can help alleviate the selection bias to make more accurate and robust predictions. The novel results of behavioral differences between users who only spread fake news and who only spread true news enable us to develop more effective tools and techniques for detecting fake news at scale. We also study the causal relationships between user profile attributes and susceptibility to spread fake news. By incorporating unbiased embeddings of fake news sharing behavior, which can fully capture confounding between user attributes and susceptibility, the causal regression model presents better performance in predicting user susceptibility. The identified causal attributes show that \textit{verified, statuses count, friends count, and org} relate significantly with user susceptibility to share fake news. This mirrors findings in psychology and social science and warrant future research for investigation of more intrinsic user attributes. 

The results here are not without limitations: as with other studies relying on social media data, there are inherently more serious issues on selection bias that our proposed model may not be able to tackle, e.g., the selection bias in the various types of friends, differences between platforms. There is also selection bias in news that is geo-located as well as language use by the individuals on different social media platforms. It is imperative to not take these data sets as being representative of the users that may be included in the datasets. Our models are also hindered by the necessary causal inference assumptions that may be violated in practice. For instance, other unmeasured confounders (e.g., categories of social media platforms) can exist in addition to the inferred fake news sharing behavior. We do not consider other important information sources such as social networks and comments of each news. The news content and attributes have yet to be fully explored. Evaluation can be further improved via interdisciplinary collaborations to obtain the ground-truth causal user attributes.  
\section*{Ethics Statement}
This work aims to advance collaborative research efforts in understanding why people spread fake news, a topic which has yet to be properly studied. Here, we provide preliminary solutions, but much work remains to bring to light the underlying causal mechanism. With our work, we hope to bring to the forefront concerns and broaden the discussions about the potential research directions in fake news. While all data used in this study are publicly available, we are committed to securing user privacy. We automatically replace user names with ordered indices in our analysis.
\section*{Acknowledgements}
This material is based upon work supported by, or in part by,
the U.S. Office of Naval Research and the U.S. Army Research Office under contract/grant number N00014-21-1-4002
and W911NF2020124. Kai Shu is supported by the John S. and James L. Knight Foundation through a grant to the Institute for Data, Democracy \& Politics at The George Washington University.
\bibliographystyle{ACM-Reference-Format}
\bibliography{sample-base}

\appendix
\section{Proof of the Proposition}
\textit{Proof.}
\begin{equation}
    \begin{aligned}
    \mathbb{E}\big[\hat{\mathcal{L}}_{unbiased}\big(\hat{\mat{S}}\big)\big]&=\mathbb{E}\Bigg[\frac{1}{|\mathcal{D}_{pair}|}\sum_{(u,i,j)\in\mathcal{D}_{pair}}\frac{Y_{u,i}}{\theta_{u,i}}\big(1-\frac{Y_{u,j}}{\theta_{u,j}}\big)\ell\big(\hat{\mat{S}}_{uij}\big)\Bigg]\\
    &=\frac{1}{|\mathcal{D}_{pair}|}\sum_{(u,i,j)\in\mathcal{D}_{pair}}\frac{\mathbb{E}[Y_{u,i}]}{\theta_{u,i}}\big(1-\frac{\mathbb{E}[Y_{u,j}]}{\theta_{u,j}}\big)\ell\big(\hat{\mat{S}}_{uij}\big)\\
    &=\frac{1}{|\mathcal{D}_{pair}|}\sum_{(u,i,j)\in\mathcal{D}_{pair}}\gamma_{u,i}\big(1-\gamma_{u,j}\big)\ell\big(\hat{\mat{S}}_{uij}\big)\\
    &=\mathcal{L}_{ideal}(\hat{\mat{S}}).
    \end{aligned}
\end{equation}
\section{Reproducibility}
\subsection{Recall@K and NDCG@K in Fake News Dissemination}
Recall@K is defined as 
\begin{equation}
\begin{aligned}
   \text{Recall@K}=\frac{\text{\# interesting fake news @K}}{\text{Total \# interesting fake news}}.
\end{aligned}
\end{equation}
NDCG@K is the normalized Discounted Cumulative Gain (DCG) defined as 
\begin{equation}
    \text{NDCG@K}=\frac{\text{DCG@K}}{IDCG@K}.
\end{equation}
We define DCG@K and IDCG@K in fake news dissemination below:
\begin{equation}
\begin{aligned}
   \text{DCG@K}=\sum_{i=1}^K\frac{2^{interest_i}-1}{\log_2(i+1)},
\end{aligned}
\end{equation}
where $interest_i$ is the interestingness of the fake news piece at index $i$, that is, $interest_i=1$ if a user is interested in this fake news piece and 0 otherwise. IDCG@K is the best possible value for DCG@K, i.e., the value of DCG for the best possible ranking of interesting fake news pieces at threshold $K$:
\begin{equation}
\begin{aligned}
   \text{IDCG@K}=\sum_{i=1}^{\text{\parbox{2cm}{interesting fake news pieces at $k$}}}\frac{2^{interest_i}-1}{\log_2(i+1)}.
\end{aligned}
\end{equation}
\subsection{Implementation}
In this section, we provide more details of the experimental setting
and configuration for reproducibility purpose.

Our proposed models were implemented in Python library Tensorflow \cite{abadi2016tensorflow} and Statsmodel \cite{seabold2010statsmodels} based on two standard models for recommender system -- BPRMF and NCF -- as we described in Sec. 5.1. The implementation code is available at: \url{https://github.com/GitHubLuCheng/Causal-Understanding-of-Fake-News-Dissemination}. For each standard model, we have three debiased models corresponding to the three proposed propensity estimates. The file names are the combinations of the recommendation model and propensity estimates, e.g., \textit{NCF\_t.py} is the code for NCF model with news-based propensity. Implementation code for baselines is adapted from \url{https://github.com/xiangwang1223/neural_graph_collaborative_filtering}.

We used publicly available datasets for fake news, FakeNewsNet \cite{shu2017fake}, available at \url{https://github.com/KaiDMML/FakeNewsNet}. For the news content, we extracted Bag of Words as features and conducted topic modeling method Latent Dirichlet Allocation (LDA) using Python package scikit-learn\footnote{\url{http://scikit-learn.org/stable/modules/generated/sklearn.decomposition.LatentDirichletAllocation.html}}. The extracted latent topics were then used as the input of the neural-network-based propensity estimates in Eq. \ref{p_neural}. We detail the parameter settings for the proposed models in Table \ref{parameters}. The descriptions of the major parameters are introduced below:
\begin{itemize}
    \item Emed\_Size: the dimensions of user and news embeddings. 
    \item Layer\_Size: the output sizes of every layer.
    \item n\_layers: the number of hidden layers. 
    \item $\lambda$: the hyperparameter for $\ell_2$ regularization.
    \item Node\_Dropout: the keep probability w.r.t. node dropout for each deep layer.
    \item Mess\_Dropout: the keep probability w.r.t. message dropout for each deep layer.
    \item Vocabulary\_Size: the threshold to control the maximum size
of vocabulary.
    \item n\_components: the number of topics.
    \item max\_df: when building the vocabulary ignore terms that have a document frequency strictly higher than the given threshold.
    \item min\_df: when building the vocabulary ignore terms that have a document frequency strictly lower than the given threshold.
\end{itemize}
\begin{table}
\caption{Details of the parameter settings in proposed models.}
\begin{tabular}{|l|l|l|}
\hline
Parameter              & BPRMF\_based model & NCF\_based model \\ \hline\hline
Epoch                  & 500          & 500        \\ \hline
Emed\_Size             & 64           & 64         \\ \hline
Layer\_Size            & 64           & 64         \\ \hline
Batch\_Size            & 1024         & 1024       \\ \hline
n\_layers              & 1            & 1          \\ \hline
$\lambda$ & 1e-2         & 1e-3       \\ \hline
Learning\_Rate         & 1e-3         & 1e-2       \\ \hline
Node\_Dropout          & 0.1          & 0.1        \\ \hline
Mess\_Dropout          & 0.1          & 0.1        \\ \hline
Vocabulary\_Size       & 2000          & 2000        \\ \hline
n\_components        & 5              &5           \\ \hline
max\_df                & 0.5           &0.5          \\ \hline
min\_df                & 5             &5            \\ \hline
\end{tabular}
\label{parameters}
\end{table}
\end{document}